\documentclass[twocolumn,aps,prd,showpacs,nofootinbib,superscriptaddress,%
preprintnumbers,floatfix,10pt]{revtex4-1}

\usepackage{amsmath}
\usepackage{amsfonts}
\usepackage{amssymb}
\usepackage{epsfig}
\usepackage{graphicx}
\usepackage{color}
\usepackage{slashed}
\usepackage{dsfont}
\usepackage{bbold}
\usepackage{mathtools}
\usepackage{array}
\usepackage{verbatim}

\newcommand{\PreserveBackslash}[1]{\let\temp=\\#1\let\\=\temp}
\newcolumntype{C}[1]{>{\PreserveBackslash\centering}p{#1}}

\usepackage{hyperref}

\newcommand{\Eqref}[1]{Eq.~\eqref{#1}}

\newcommand{\Nf}{N_{\mathrm{f}}}
\newcommand{\Np}{N_{\mathrm{p}}}
\newcommand{\pat}{\partial_t}

\DeclareMathOperator{\SU}{SU}

\newcommand{\g}{\gamma} 
\newcommand{\barh}{\bar{h}}

\begin{document}

\preprint{}

\title{Global effective-potential flow of a self-organized critical relativistic Luttinger-Yukawa model}

\author{Tim Eckert}
\email{tim.eckert@uni-jena.de}
\affiliation{Theoretisch-Physikalisches Institut,
Abbe Center of Photonics, Friedrich Schiller University Jena, Max Wien
Platz 1, 07743 Jena, Germany}
\author{Holger Gies}
\email{holger.gies@uni-jena.de}
\affiliation{Theoretisch-Physikalisches Institut, 
Abbe Center of Photonics, Friedrich Schiller University Jena, Max Wien 
Platz 1, 07743 Jena, Germany}
\affiliation{Helmholtz-Institut Jena, Fröbelstieg 3, D-07743 Jena, Germany}
\affiliation{GSI Helmholtzzentrum für Schwerionenforschung, Planckstr. 1, 
64291 Darmstadt, Germany} 

\begin{abstract}
We revisit a recently proposed Yukawa model with relativistic Luttinger fermions
which features self-organized criticality: for generic initial conditions, the
renormalization flow approaches an infrared-attractive partial fixed point at
which all couplings are irrelevant except for a marginal scalar mass parameter.
This parameter drives the system towards a symmetry-broken phase logarithmically
slowly. Using the functional renormalization group, we resolve the full field
dependence of the effective potential, overcoming the small convergence radius
of previously used local expansions. In the large-$N_\text{f}$ limit, we find a
global analytic solution of the partial fixed-point equation, which exhibits
Coleman-Weinberg-type large-field asymptotics that persists at higher-orders in
$1/N_\text{f}$. The solution agrees well with numerical results, even for
$N_\text{f}=1$. By numerically integrating the full flow from the symmetric into
the symmetry-broken regime, we determine the universal mass ratios of the model
as functions of $N_\text{f}$ improving upon previous estimates in particular
for small $N_\text{f}$.
\end{abstract}

\pacs{}

\maketitle

\section{Introduction}
\label{intro}

Self-organized criticality \cite{Bak:1987xua,Bak:1988zz} is a property of a
variety of systems in statistical physics and is characterized by the emergence of
scale invariance without the need for fine-tuning of parameters. It occurs in
slowly driven nonequilibrium many-body systems with nonlinear dynamics
\cite{Manna:1991,Olami:1992,Malamud:1998,Vespignani:1998}, or in stochastic
growth models \cite{Kardar:1986xt,Canet:2009vz,Canet:2025sla} to name a few
examples. In all these settings, the critical point can be associated with a
fixed point of the renormalization group (RG) flow, which is approached
dynamically and largely independently of the initial conditions. Self-organized
criticality is also intensely discussed in astrophysical contexts such as solar
flares \cite{Lu:1991,Aschwanden:2014dna}. 

Recently, features of self-organized criticality have been observed in a
relativistic Yukawa theory with Luttinger fermions \cite{Gies:2025pqv}. These
features manifest themselves in rather different renormalization flows compared
to conventional models. In fact for generic initial conditions, all coupling
terms studied so far are found to be renormalization-group irrelevant, and
the system is driven to a partial fixed point towards the IR. This partial fixed
point renders the system approximately scale invariant over a wide range of RG
scales.

In contradistinction to the examples of statistical physics, the partial fixed
point of the model of Ref.~\cite{Gies:2025pqv} is ultimately destroyed by a
marginal direction in the RG flow which is associated with a scalar mass parameter
that runs to negative values and thus triggers a symmetry breaking transition of
the flow regime. In the long-range limit, the model hence ends up in the broken phase --
again independently of the initial conditions -- and is characterized by a
massive spectrum of excitations.  Provided that the RG flow has spent a long RG
time in the vicinity of the partial fixed point, the resulting mass spectrum is
exponentially small compared to the initial UV scale and their ratios are a universal prediction
of the model. 

From the perspective of relativistic quantum field theory, this model partly
realizes a scenario developed in \cite{Bornholdt:1992up} in an attempt to
address the naturalness, or hierarchy problem of the standard model of particle
physics; see also \cite{Eroncel:2018dkg} for a concrete model scenario involving
degrees of freedom beyond the standard model. 

Many of the examples of self-organized criticality mentioned above have in
common that a quantitative understanding requires sophisticated methods or
large-scale numerical simulations. From the RG viewpoint, a control of the
system near the fixed point is needed; since it is generically a non-Gaussian
fixed point, perturbative weak-coupling methods are not sufficient. For
instance, for the Yukawa model of \cite{Gies:2025pqv}, the partial fixed point
becomes visible in a large-$\Nf$ expansion, where $\Nf$ is the number of fermion
flavors. Further quantitative results have been obtained using the functional RG
in a derivative expansion together with a polynomial expansion of the effective
potential. 

The present work is devoted to a more robust analysis of the RG flow of the
Yukawa model of \cite{Gies:2025pqv}. We provide both analytical and numerical
results for the RG flow of the full effective potential and further running
couplings. In fact from a method perspective, the application of the functional
RG to self-organized criticality in the sense of the present relativistic model
is a challenge. This is because a numerical code must be able to (i) address
the flow towards and in the vicinity of the partial fixed point typically
involving many orders of magnitude in the RG scale, and (ii) resolve the flow
towards the symmetry-broken phase, which is eventually triggered by the
marginally relevant scalar mass parameter. The latter requires a careful
treatment of the effective potential in the broken phase including the approach
to convexity \cite{Tetradis:1992qt,Litim:2006nn}. While (i) represents a slow logarithmic flow, (ii) is
characterized by a rapid flow towards the IR including the build-up of
non-analyticities. In the present work, we demonstrate that both aspects can be
addressed with suitable numerical methods, yielding quantitative results for the
universal long-range observables. 

In Sect.~\ref{sec:model}, we summarize basic aspects of the model and its RG
flow in a derivative expansion. Section~\ref{sec:AR} presents novel analytical
results for the RG flow of the effective potential in the vicinity of the
partial fixed point. In Sect.~\ref{sec:Numerical}, we present our numerical
results for the RG flow including a comparison to the analytical results. We
also present our quantitative best estimates for the universal long-range
observables of the model in the present approximation scheme. We conclude in
Sect.~\ref{sec:conc}.

\section{Renormalization flow of the $\g_{11}$ Yukawa model}
\label{sec:model}

\subsection{Model and symmetries}

Our focus is on a Yukawa model describing the interaction of a real scalar field
$\phi$ with relativistic Luttinger fermions $\psi$ in $d=4$ spacetime
dimensions. The model first appeared in a mean-field analysis of a purely
fermionic model in \cite{Gies:2024dly}, and was subsequently studied in a
functional RG approach in \cite{Gies:2025pqv}. The Euclidean action of the model
is given by
\begin{eqnarray}
    S&=& \int d^4x \left[  - \bar\psi G_{\mu\nu}\partial^\mu \partial^\nu \psi 
    + \frac{1}{2} \partial_\mu \phi  \partial^\mu \phi \right. \nonumber\\
    && \left. \quad \qquad - \bar{h} \phi \bar\psi \gamma_{11} \psi + \frac{\bar{m}^2}{2} \phi^2 
    + \frac{\bar{\lambda}}{8} \phi^4 \right]. 
    \label{eq:S11}
\end{eqnarray}
Here, $\psi$ denotes a Luttinger spinor with $d_\gamma = 32$ components, and
$\bar{\psi}$ its conjugate $\bar{\psi} = \psi^\dagger h$, where $h$ is a spin
metric. For each spacetime index $\mu,\nu$, the $d_\gamma \times d_\gamma$
matrices $G_{\mu\nu}$ are tracefree in spinor space, symmetric with respect to
their spacetime indices $G_{\mu\nu}=G_{\nu\mu}$, and satisfy the relativistic
version of the (Euclidean) Abrikosov algebra \cite{Abrikosov:1974a},
\begin{equation}
    \{G_{\mu\nu},G_{\kappa\lambda}\} = \frac{-2}{d-1} \delta_{\mu\nu} \delta_{\kappa\lambda}
     + \frac{d}{d-1} \left( \delta_{\mu\kappa} \delta_{\nu\lambda} + \delta_{\mu\lambda} 
     \delta_{\nu\kappa} \right) \mathbb{1}_{d_\gamma},
    \label{eq:AA}
\end{equation}
for general spacetime dimension $d$. In addition, the matrices $G_{\mu\nu}$ are
tracefree with respect to their spacetime indices, $G^{\mu\nu} \delta_{\mu\nu} =
0$. The Abrikosov matrices $G_{\mu\nu}$ can be constructed from linear
combinations of a suitable set of $d_\gamma \times d_\gamma$ (Euclidean) Dirac
matrices $\gamma_A$, satisfying the Clifford algebra $\{\gamma_A,\gamma_B\} = 2
\delta_{AB} \mathbb{1}_{d_\gamma}$. Together with the spin metric $h$, the nine
different Abrikosov matrices can be spanned by $A=1,\dots,10$ Dirac matrices,
see \cite{Gies:2024dly} for a concrete representation. The Dirac-Clifford
algebra features an additional matrix $\gamma_{11}$, which anticommutes with all
$\gamma_A$ and squares to the identity. This matrix is used to construct the
Yukawa interaction term in Eq.~\eqref{eq:S11}.

The spinor part of the model inherits a global SL(32,$\mathbb{C})$ spin-base
symmetry from the similarity transformations of the Euclidean Dirac algebra. The
Lorentz invariance of the model in Minkowski space can be understood as a
subgroup of this spin-base symmetry. In comparison to non-relativistic models
with Luttinger fermions
\cite{LuttingerPhysRev.102.1030,Murakami:2004zz,Moon:2012rx,Janssen:2015xga,Janssen:2016xvc,Ray:2018gtp,Dey:2022lkx,Moser:2024dmq},
the spin metric $h$ is a novel ingredient ($h= \mathds{1}$ is a legitimate
choice in the non-relativistic case). Correspondingly, the irreducible
representation for the relativistic case can be twice as large as for the
non-relativistic case, i.e., $d_\gamma = 32$ for $d=4$, whereas the construction
of the $G_{\mu\nu}$ can follow the same procedure as in the non-relativistic
case \cite{Janssen:2015xga} apart from the necessary (anti-)hermiticity
properties in Minkowski space.  

The model features a discrete $\mathbb{Z}_2$ symmetry under which
$\psi\to-e^{i\frac{\pi}{2} \g_{11}} \psi$, $\bar\psi \to \bar\psi
e^{-i\frac{\pi}{2} \g_{11}}$, and $\phi\to-\phi$ (with $\psi$ and $\bar\psi$
being independent variables in the Euclidean quantum theory). This symmetry is
reminiscent of a discrete chiral symmetry known from the corresponding
Dirac-Yukawa model \cite{Hofling:2002hj,Holland:2003jr,Fodor:2007fn,Braun:2010tt,Mesterhazy:2012ei,Gies:2013fua,Vacca:2015nta,Jakovac:2015kka,Gies:2017ajd}. Analogously, the 
spontaneous breaking of this symmetry is accompanied by mass gap generation: while
a nonzero vacuum expectation value $v$ of the scalar field goes along with a
standard $\sigma$-type mass for excitations of the scalar field, the Yukawa term
generates a pair of complex conjugate mass gaps for the fermionic modes
\cite{Gies:2024dly}. Nevertheless, the mass gaps arrange themselves such that
the RG flow freezes out in the deep IR.

Generalizing the model to $\Nf$ flavors of Luttinger fermions, the model
features a global $\SU(\Nf)$ flavor symmetry which is preserved by the Yukawa
interaction and remains intact also in the $\mathbb{Z}_2$-broken phase. The
flavor symmetry is not essential for the self-organized criticality of the
model, but it allows for a systematic $1/\Nf$ expansion.

\subsection{Functional RG flow}

Let us briefly recall the functional RG flow of the model, for more details we
refer the reader to \cite{Gies:2025pqv}. Using the Wetterich equation
\cite{Wetterich:1992yh}, we are interested in the flow of the scale-dependent
effective action $\Gamma_k$ that interpolates between the microscopic action $S$
at $k=\Lambda$ (a UV scale) and the quantum effective action $\Gamma$ at $k=0$.
We solve the Wetterich equation using the ansatz
\begin{eqnarray}
    \Gamma_k &=& \int d^4x \left[  - Z_\psi \bar\psi G_{\mu\nu}\partial^\mu \partial^\nu \psi 
    + \frac{Z_\phi}{2} \partial_\mu \phi  \partial^\mu \phi \right. \nonumber\\
    && \left. \quad \qquad - \bar{h} \phi \bar\psi \gamma_{11} \psi + U(\bar\rho) \right], 
    \quad \bar\rho= \frac{1}{2} \phi^2,
    \label{eq:Gammak}
\end{eqnarray}
where the wave function renormalizations $Z_{\psi,\phi}$, the Yukawa coupling
$\bar{h}$, and the effective potential $U(\bar\rho)$ are considered to depend on the RG
scale $k$. The ansatz \eqref{eq:Gammak} can be considered as an improved
leading-order derivative expansion, i.e., the analogue of the improved
local-potential approximation (LPA$^\prime$) for a purely scalar theory
\cite{Morris:1993qb,Morris:1998da,Bagnuls:2000ae,Berges:2000ew,Delamotte:2007pf}. The ansatz has proved useful in corresponding
Dirac-Yukawa cases \cite{Jungnickel:1995fp,
Hofling:2002hj,Diehl:2009ma,Braun:2010tt,Mesterhazy:2012ei,Jakovac:2014lqa,
Janssen:2014gea,Vacca:2015nta,Gies:2014xha,Classen:2015mar,Knorr:2016sfs,Fu:2016tey,
Stoll:2021ori,Gies:2023jzd}
and can be systematically improved by including higher-order derivative terms as
well as considering $Z_{\psi,\phi}$, $\bar{h}$, and higher-order couplings also as
functions of $\bar\rho$, see \cite{Balog:2019rrg,Knorr:2016sfs,Dupuis:2020fhh}.   

The Wetterich equation provides us with RG flow equations for the
scale-dependent couplings and the effective potential. For the analysis of the
partial fixed point, we use dimensionless renormalized quantities such as the
dimensionless potential and the dimensionless Yukawa coupling,
\begin{equation}
    u(\rho)= \frac{U(\bar\rho)}{k^d}, \quad 
    \rho = \frac{Z_\phi}{2} \frac{\phi^2}{k^{d-2}}; \qquad   h^2=\frac{\barh^2}{Z_\psi^2 Z_\phi k^{6-d}}.
    \label{eq:dimless}
\end{equation}
The flow of the wave function renormalizations is summarized in the anomalous
dimensions
\begin{equation}
    \eta_{\psi,\phi}= - \pat \ln Z_{\psi,\phi}. \label{eq:etas}
\end{equation}
Following \cite{Gies:2025pqv}, the resulting flow of the effective potential reads
\begin{eqnarray}
    \pat u(\rho) & =&  -d \, u + (d-2+\eta_\phi) \rho u' \nonumber\\
    &&+ 2 v_{d}\Big[ l_{0}^{d}\left(u' + 2\rho  u''; \eta_\phi \right) \nonumber\\
    &&\qquad -\Nf d_\g  l_{0}^{(\text{L})\,d}\left(2 \rho h^2 ; \eta_\psi \right) \Big],
  \label{eq:uflow} 
\end{eqnarray}
where $\pat= k \frac{d}{dk}$ and $v_d^{-1} = 2^{d+1} \pi^{d/2} \Gamma(d/2)$.
Primes denote derivatives with respect to $\rho$. The threshold functions
$l_{\dots}^{\dots}$ encode the decoupling of massive modes and
depend on the choice of the regularization; they approach positive constants for small
arguments and decay to zero for large arguments. Their explicit form for the
partially linear regulator \cite{Litim:2001up} is given in
App.~\ref{app:thresh}.  The flow of the Yukawa coupling can be written as
\begin{eqnarray}
    \pat h^2&=& -(6-d-2\eta_\psi -\eta_\phi) h^2 \nonumber\\
    &&+ 8v_d \, h^{4} 
    l_{1,1}^{\text{(LB)}\,d}\left(\omega_{1},\omega_{2};
         \eta_{\psi},\eta_{\phi}\right),  \label{eq:flowhq}\\
&& \text{with}\,\,\, \omega_1  =  2 \, \kappa \, h^2, \,\,
\omega_2  =  u'(\kappa) + 2 \kappa u''(\kappa)
. \label{eq:omegadef}  
\end{eqnarray}
Here, $\kappa$ denotes the location of the minimum of the effective potential,
i.e., $u'(\kappa)=0$ for $\kappa\neq 0$. The anomalous dimensions satisfy
\begin{eqnarray}
\eta_\psi\!&=& \frac{16}{d(d+2)} v_d h^2 m_{1,2}^{(\text{LB}),d}(\omega_1,\omega_2;\eta_\psi,\eta_\phi), \label{eq:etapsi}\\
\eta_\phi\!&=& \frac{8v_d}{d} \kappa (3u'' +2\kappa u''')^2 m_{2,2}^{d}(\omega_2;\eta_\phi) \nonumber\\
&&\!\!+\frac{8v_d}{d}\Nf d_\gamma h^2 \left(m_4^{(\text{L}),d}(\omega_1,\eta_\psi) -2 h^2\kappa m_2^{(\text{L}),d}(\omega_1,\eta_\psi)\! \right), \nonumber\\
&& \label{eq:etaphi}
\end{eqnarray}
where the $m_{\dots}^{\dots}$ denote further threshold functions, cf. App.~\ref{app:thresh}.
For the partially linear regulator, the $\eta_{\psi,\phi}$ equations are algebraic and can be solved in closed form. Inserting them into the flow equations for $u$ and $h^2$ yields a partial differential equation for $u$ and an ordinary differential equation for $h^2$. 

In the following, let us summarize a few results for $d=4$ from \cite{Gies:2025pqv} that
are relevant for the present work. These results are based on a polynomial
expansion of the effective potential of the forms,
\begin{equation}
u(\rho)\simeq \left\{ 
    \begin{aligned}
        &\sum_{n=1}^{\Np} u_n \rho^n, \quad \text{(SYM)},\\
        &\sum_{n=2}^{\Np} u_n (\rho-\kappa)^n, \quad \text{(SSB)},
    \end{aligned}
\right.
\label{eq:upolexp}
\end{equation}
depending on whether the minimum of the potential is in the symmetric regime
(SYM) at $\kappa=0$ or in the spontaneously symmetry-broken regime (SSB) at
$\kappa\neq 0$. The expansion is truncated at order $\Np$ and the flow equations
for the couplings $u_n$ are obtained from a corresponding expansion of
\eqref{eq:uflow}.

The partial fixed point of the model becomes visible for generic initial
conditions in the SYM regime: for $\kappa=0$, the scalar anomalous dimension in
\eqref{eq:etaphi} is manifestly positive and scales as $\eta_\phi \sim \Nf
d_\gamma h^2$. Since $\Nf d_\gamma \geq 32$ is large even for $\Nf=1$, the
anomalous dimension dominates the flow of the Yukawa coupling in
\eqref{eq:flowhq} competing with the canonical scaling term, cf.
Eq.~\eqref{eq:flowhqNf} below. This induces a fixed point for the Yukawa
coupling at a finite value $h^2_\ast$ which is IR attractive (RG irrelevant), 
and at which the scalar anomalous dimension takes a finite value
$\eta_\phi^\ast\simeq 2$. Due to
small corrections from $\eta_\psi$ and the loop term $\sim h^4$ in
\eqref{eq:flowhq}, the fixed-point value of $\eta_\phi^\ast$ is slightly smaller
than 2 and approaches 2 in the limit $\Nf\to\infty$. 

With $\eta_\phi^\ast\simeq 2$, all couplings $u_{n\geq 2}$ become RG irrelevant,
exhibiting non-Gaussian IR fixed points which are manifestations of the partial
fixed point of the functional RG flow of all interactions. The only remaining
coupling is the mass parameter $u_1=\epsilon$ that satisfies an approximate flow
equation of the form
\begin{equation}
    \pat \epsilon \simeq c_\epsilon, 
    \label{eq:floweps}
\end{equation}
where $c_\epsilon$ at the partial fixed point depends only weakly on $\epsilon$
itself and approaches $c_\epsilon\to 4/5$ for large $\Nf$. Therefore, the flow
of the mass parameter does not exhibit a fixed point -- being the reason why we
talk about a partial fixed point for all other couplings. Nevertheless, the mass
parameter obeying \eqref{eq:floweps} runs only logarithmically with the RG scale
such that the system can spend a wide range of scales near the partial fixed
point for a wide range of initial conditions without the need for fine-tuning.
It is this feature that we associate with self-organized criticality in the
present relativistic Yukawa model.

Since the mass parameter is marginally relevant, the system ultimately runs into
the SSB regime with $\epsilon<0$ and thus $\kappa>0$. Here the anomalous
dimension $\eta_\phi$ rapidly decreases to small values, the system departs from
the partial fixed point, and, after a short RG time, the flow freezes out in the
deep IR with a massive spectrum of excitations. 

The long-range observables of the $d=4$ dimensional model studied in \cite{Gies:2025pqv} include
the vacuum expectation value $v$, the scalar $\sigma$-type mass $m_\sigma$ and
the fermionic mass gap $m_\psi$ obtained from the IR flow via 
\begin{equation}
    v= k \sqrt{2\kappa}|_{k\to0},\,\, 
    m_\sigma=k \sqrt{2\lambda\kappa}|_{k\to0},\,\, 
    m_\psi = k \sqrt[4]{2\kappa h^2}|_{k\to0},
    \label{eq:obs}
\end{equation} 
where $\lambda = u''(\kappa)$ is the quartic scalar coupling at the 
minimum of the potential.
Provided that the RG flow has spent a long RG time in the vicinity of the
partial fixed point, all these dimensionful observables are exponentially small
compared to the initial UV scale $\Lambda$ and their ratios approach universal values. In particular,
estimates for the universal ratios $m_{\sigma,\psi}/v$ and $m_\sigma/m_\psi$
have been obtained in \cite{Gies:2025pqv} using a polynomial expansion of the
effective potential up to order $\Np=22$. While the results for the universal
ratios show a reasonable apparent convergence with increasing $\Np$, the IR
potential itself has been resolved only in a very small neighborhood of the
minimum $\kappa$; no global information in field space has been obtained. 

We conclude this review of previous results by comparing the $\gamma_{11}$
Yukawa model to a corresponding Dirac-Yukawa model. In the latter, the Yukawa
interaction and the $\phi^4$ interaction are both marginal near the Gaussian
fixed point in $d=4$ dimensions while the scalar mass parameter is relevant. In
the SSB phase, the scalar and fermion mass are fixed by the initial conditions
of the interactions, whereas the vacuum expectation value can be viewed as fixed
by the initial condition of the scalar mass parameter. In order to obtain a
large hierarchy between the vacuum expectation value $v$ and the UV scale
$\Lambda$, the initial condition of the scalar mass parameter must be fine-tuned
rather precisely to the critical value separating the symmetric and the SSB
phase -- rendering a large scale separation unnatural. By contrast, the
$\gamma_{11}$ Yukawa model is always in the SSB phase and a large scale
separation is natural for generic initial conditions that allow the system to
spend a long RG time in the vicinity of the partial fixed point. 

Another difference is that the fermion mass parameter $m_\psi$ generated by the
$\gamma_{11}$ Yukawa model gaps the spectrum of the Luttinger fermions along the
imaginary axis in the $p^2$ plane \cite{Gies:2024dly}.

\section{Global analysis of the partial fixed point}
\label{sec:AR}

It is one of the powerful features of the functional RG that it allows for a
global analysis of the effective potential in field amplitude space. The fact
that the flow equation for the potential \eqref{eq:uflow} is a partial
differential equation can give access to a conceptual analysis of the parameter
space of the model. In fact, the criterion of global existence of solutions can
reduce the space of admissible parameters and single out physical solutions
\cite{Hasenfratz:1985dm,Morris:1994ki,Codello:2012ec,
Hellwig:2015woa, Borchardt:2015rxa, Dietz:2015owa,Demmel:2015oqa,Litim:2016hlb,Juttner:2017cpr,Gies:2015lia,Ohta:2015efa,Gies:2016kkk,Gies:2018vwk,Gies:2019nij,Knorr:2016sfs,Dabelow:2019sty,Schreyer:2020,Gies:2024xzy}.
Even more so, global existence can also reveal artifacts of approximations, such
as the polynomial expansion \cite{Laporte:2022ziz,deBrito:2023myf}. In view of the small radius
of convergence for the polynomial expansion found in \cite{Gies:2025pqv} for the
present model, a global analysis is particularly important. 
For the remainder of this paper, we mostly consider $d = 4$ dimensions.

\subsection{Large-$N_\text{f}$ Limit}

Let us start with an analysis of the flow equations in the large-$N_\text{f}$
limit, using two assumptions that turn out to be self-consistent: (i) at a fixed
point or a partial fixed point, we expect the anomalous dimensions in
Eqs.~\ref{eq:etaphi} and \ref{eq:etapsi} to approach constants; (ii) in line
with the findings of \cite{Gies:2025pqv}, we assume that the minimum of the
effective potential is located in the symmetric regime for the scales of
interest, i.e., $\kappa=0$. 

From \Eqref{eq:etaphi}, we then can deduce that the first line vanishes and the
second line implies that the Yukawa coupling at a (partial) fixed point scales
as $h^2 \sim 1/N_\text{f}$ for large $N_\text{f}$. Upon insertion into
\Eqref{eq:etapsi}, we find that the fermionic anomalous dimension $\eta_\psi$
approaches zero in this limit. With this, the flow of the Yukawa coupling in
\Eqref{eq:flowhq} simplifies to
\begin{equation}
    \partial_t h^2 = -\left( 2 - \eta_\phi \right) h^2 \,.
    \label{eq:flowhqNf}
\end{equation}
Since $\eta_\phi \sim h^2$ for large $\Nf$, the flow of $h^2$ exhibits a
nontrivial (partial) fixed point at $h^2=h_\ast^2>0$ where
$\eta_\phi^\ast = 2$, cf. \Eqref{eq:etaphi}. This demonstrates
that assumption (i) is self-consistent, provided that also the other sectors
exhibit a (partial) fixed point. 

For the analysis of the scalar sector, it is useful to study the flow of the
derivative of the potential with respect to the invariant $\rho$. Defining
$y(\rho)= u'(\rho)$, we obtain from \Eqref{eq:uflow} 
\begin{equation}
\begin{aligned}
    \partial_t y = {} &(-2+\eta_\phi) y+ (2+\eta_\phi) \rho y' \\
    &- v_4 \left[\left(1-\frac{\eta_\phi}{6} 
    \right)\frac{3 y'+2 \rho y''}{(1+y+2 \rho y')^2}\right. \\
    &\qquad\quad \left. 
        - 2 N_\text{f} d_\gamma \left(1-\frac{\eta_\psi}{6} \right)
        \frac{2 h^2}{(1+2 \rho h^2)^2}\right]\, ,
    \label{flowequd}
\end{aligned}
\end{equation}
where $v_4=\frac{1}{32 \pi^2}$. For simplicity, we have used the partially
linear regulator \cite{Litim:2001up} adapted to the present model, see
App.~\ref{app:thresh}. For this regulator, the partial fixed point of the
Yukawa coupling is located at $h_\ast^2 = \frac{32 \pi^2}{5\Nf d_\gamma}$.

Now, it is instructive to evaluate the flow equation \eqref{flowequd} at
$\rho=0$ for large $N_\text{f}$. In agreement with \cite{Gies:2025pqv}, we find
for the dimensionless renormalized mass parameter $\epsilon = y(\rho=0)=
\frac{\bar{m}^2}{Z_\phi k^2}$,
\begin{equation}
\partial_t \epsilon = c_\epsilon \simeq \frac{4}{5} + O(1/N_\text{f}) \, .
\label{eq:flowepsNf}
\end{equation}
The fact that the flow of $\epsilon$ is given by a positive constant, tells us
that the mass parameter is marginally relevant. For positive initial values
$\epsilon_\Lambda$ at the UV scale $\Lambda$, $\epsilon= \epsilon_\Lambda +
c_\epsilon \ln(k/\Lambda)$ runs logarithmically slowly towards smaller values
and eventually becomes negative in the IR. In absence of any further minima of
the effective potential, the fact that $\epsilon>0$ holds for a wide range of
scales demonstrates that also assumption (ii) is self-consistent in this regime. 

Continuing the polynomial expansion of the potential to higher orders, all
higher-order couplings $u_{n\geq 2}$ are found to exhibit non-Gaussian fixed
points at which they are RG irrelevant \cite{Gies:2025pqv}. This is the partial
fixed point manifesting self-organized criticality. In order to overcome the
limitations of the polynomial expansion, specifically the small radius of
convergence, we add a third assumption that also turns out to be
self-consistent: (iii) the flow of the effective potential near the partial
fixed point is quantitatively dominated by its marginal direction, i.e.,
$\partial_t y(\rho) \simeq \partial_t \epsilon \simeq 4/5$ for all $\rho$. 

At first sight, this assumption seems justified only in a vicinity of $\rho=0$.
However, studying the scaling of all terms on the right-hand side of
\Eqref{flowequd} with respect to $N_\text{f}$, such that the left-hand side is
of order $O(1)$, we deduce
\begin{equation}
    u \sim N_\text{f}, \quad \text{and} \quad \rho \sim N_\text{f}.
    \label{Nfsc}
\end{equation}
This suggests that assumption (iii) could also hold for large $\rho$ at least in
the large-$N_\text{f}$ limit. With the replacement on the left-hand side
suggested by (iii), we can solve the resulting differential equation,
\begin{eqnarray}
\frac{4}{5} &=& (-2+\eta_\phi) y+ (2+\eta_\phi) \rho y' \nonumber\\
&&+ 2v_\text{4} N_\text{f} d_\gamma \left(1-\frac{\eta_\psi}{6} \right)
        \frac{2 h^2}{(1+2 \rho h^2)^2}\, , \label{flowequd45} 
\end{eqnarray}
even analytically for constant $h^2, \eta_\phi, \eta_\psi$. The solution reads
\begin{align}
    y(\rho) = {}&
    c\,\bigl[(2+\eta_\phi)\rho\bigr]^{\frac{2-\eta_\phi}{2+\eta_\phi}}
    + \frac{1}{15(-2+\eta_\phi)} \biggl[ 12
    \notag\\
    &\quad - 10\,d_\gamma h^2 N_\text{f} v_4 (6-\eta_\psi)
    \label{anasolNf}\\
    &\qquad \times {}_2F_1\!\left(
        2, \frac{-2+\eta_\phi}{2+\eta_\phi},
        \frac{2\eta_\phi}{2+\eta_\phi},
        -2h^2\rho
    \right) \biggr] \,,
    \notag
\end{align}
where $c \in \mathbb{R}$ is an integration constant and ${}_2F_1$ is the
hypergeometric function. With \Eqref{flowequd45} representing the partial
fixed-point equation in the large-$N_\text{f}$ limit, the solution
\eqref{anasolNf} is a global solution for the fixed-point potential in this
limit. In the large-$\rho$ limit, we obtain the large-$\rho$ asymptotics
\begin{align}
    y(\rho) = {}&
    \frac{4}{5(-2+\eta_\phi)}
    + c\bigl[(2+\eta_\phi)\rho\bigr]^{
        \frac{2-\eta_\phi}{2+\eta_\phi}}
    \notag\\
    &+ \frac{
        16^{\frac{1}{2+\eta_\phi}}\,
        d_\gamma N_\text{f} v_4 (-6+\eta_\psi)
    }{3(2+\eta_\phi)\rho}\,
    (h^2\rho)^{\frac{4}{2+\eta_\phi}}
    \notag\\
    &\qquad\times
    \Gamma\!\left(\frac{-2+\eta_\phi}{2+\eta_\phi}\right)
    \Gamma\!\left(\frac{6+\eta_\phi}{2+\eta_\phi}\right)\,.
    \label{anasolNflrho}
\end{align}
In addition to the homogeneous solution, a fermionic contribution thus
persists in the large-$\rho$ limit. It scales with the same power,
$\rho^{\frac{2-\eta_\phi}{2+\eta_\phi}}$, as the homogeneous solution. Although
this exponent vanishes as $\eta_\phi\to\eta_\phi^*=2$, the prefactor is singular
in this limit. At the partial fixed point, with $\eta_\psi^*=0$ and
$h_\ast^2=\frac{32\pi^2}{5N_\text{f}d_\gamma}$, this singularity cancels the
pole $\frac{4}{5(-2+\eta_\phi)}$ in the constant term. Taking the limit
$\eta_\phi\to\eta_\phi^*=2$ then yields the logarithmic scaling
\begin{equation}
    y(\rho) \sim \frac{1}{5}\log\rho\,.
    \label{anasolNflrho2}
\end{equation}
Alternatively, we can directly insert the large-$N_\text{f}$
fixed-point values $\eta_\phi^*=2$, $\eta_\psi^*=0$, and
$h_\ast^2=\frac{32\pi^2}{5N_\text{f}d_\gamma}$. In this limit,
the fixed-point equation~\eqref{flowequd45} simplifies to
\begin{equation}
    \frac{4}{5}
    = 4\rho\,y'
    + \frac{4}{
        5\left(1+\frac{64\pi^2\rho}{5N_\text{f}d_\gamma}\right)^2
    }\,.
\end{equation}
Its solution is
\begin{align}
    y(\rho) = {}&
    c - \frac{N_\text{f}d_\gamma}{
        5N_\text{f}d_\gamma+64\pi^2\rho}
    \notag\\
    &+ \frac{1}{5}
    \log\!\left(5N_\text{f}d_\gamma+64\pi^2\rho\right)\,.
\end{align}
The large-$\rho$ asymptotics reproduce Eq.~\eqref{anasolNflrho2},
confirming the consistency of the two approaches.

Correspondingly, the large-$\rho$ asymptotics of the effective potential
acquires a Coleman-Weinberg-type logarithmic behavior, $u(\rho) \sim \rho \log
\rho$ \cite{Coleman:1973jx}. These simplified solutions can equivalently be derived by starting from
the partial fixed point equations, where the linear term $(-2+\eta_\phi) y$
vanishes identically for $\eta_\phi = 2$.

Alternatively, this system can be analyzed by taking an additional derivative of
the flow of the effective potential with respect to $\rho$ before evaluating the
differential equation. This procedure circumvents assumption (iii) and the
corresponding approximation on the left-hand side of the flow equation. This
approach yields structurally identical differential solutions. It
introduces an additional integration constant that maps directly to the
$\rho$-independent sector of the solutions derived above.

\subsection{Small- and Large-Field Expansion}
\label{slfieldexp}

In order to go beyond the large-$N_\text{f}$ limit, our analyses here and in the
subsequent sections require a functional definition of the partial fixed point.
Such a definition is not obvious, since the flow of the effective potential is
not zero at the partial fixed point. In the small-field expansion, it is
possible to identify the partial fixed point as a fixed point of all coupling
flows, once the flow of the (marginal) mass parameter has been isolated.

On the functional level, we can define the partial fixed point as a solution of
a residual fixed-point equation, once the flow of the marginal direction in
field space has been isolated. For this, let us assume that $\epsilon$ is not
just the dimensionless mass parameter, but serves as a coordinate along the
marginal direction in field space; then the flow of the effective potential can
be written as
\begin{equation}
    \partial_t u(\rho) \equiv \partial_t u(\rho;\epsilon) 
=  \partial_t u(\rho;\epsilon) \Big|_{\epsilon} 
+ \frac{\partial u(\rho;\epsilon)}{\partial \epsilon} \, \partial_t \epsilon \, . 
\label{eq:partialFPdef}
\end{equation}
In analogy to the small-field expansion, we define the partial fixed
point by the condition that $\partial_t u(\rho;\epsilon) \Big|_{\epsilon} = 0$
for all $\rho$. This condition depends on $\epsilon$ as expected, characterizing
the drift of the partial-fixed point with the marginal direction in field space.
At the partial fixed point, we thus obtain from \Eqref{eq:partialFPdef} the
condition 
\begin{equation}
  \frac{\partial u(\rho;\epsilon)}{\partial \epsilon} \, \partial_t \epsilon 
  =  \partial_t u(\rho) \equiv\, 
  \text{\Eqref{eq:uflow}}  \, .
  \label{eq:partialFPdef2}
\end{equation}
Since the very definition of the partial fixed point is based on the assumption
of $\epsilon$ parametrizing the marginal direction (as will be confirmed by the
numerical results), we expect that $\partial_t \epsilon= c_\epsilon$ is a
constant, possibly depending on $\epsilon$ and $N_\text{f}$. Nevertheless, the
marginal direction characterized by $\frac{\partial u(\rho;\epsilon)}{\partial
\epsilon}$ in general is an a priori unknown full function in field space. For
the purpose of the present analysis, however, we try to approximate this
function by an ansatz that has sufficient overlap with the marginal direction.
From the small field expansion, we know that the marginal direction is dominated
by the mass parameter, $ \frac{\partial u(\rho;\epsilon)}{\partial \epsilon} =
\rho  + O(\rho^2)$. While higher-order terms could be worked out systematically,
we restrict ourselves to the leading-order term in the present analysis.
This is justified by the fact that perturbations about a fixed point are
required to be polynomially bounded in order to be considered as physical
perturbations \cite{Morris:1998da}. Moreover, retaining only the leading-order
term is supported by the full numerical results below.
Furthermore, confining the ansatz to the leading-order term ensures that it remains
subdominant compared to the leading-order large-field asymptotics studied in the
previous section; the latter is thus not spoiled by our ansatz. 

To specify the partial fixed-point equation, we also need an
approximation for $c_\epsilon$. The small-field expansion gives
$c_\epsilon=4/5$ at leading order in the large-$\Nf$ limit.
The perturbative analysis of Ref.~\cite{Gies:2025pqv}, including
the quartic scalar coupling, yields the following large-$\epsilon$
estimate for arbitrary $\Nf$:
\begin{equation}
    c_\epsilon \simeq \frac{4}{5}-\frac{1}{20\Nf}\, .
    \label{eq:cepsNf}
\end{equation}
The full numerical flow exhibits only a weak dependence of $c_\epsilon$ on
$\epsilon$ and $\rho$ at the partial fixed-point, see
Fig.~\ref{fig:c_epsilon_flow} below. We therefore use this estimate as a
constant approximation on the left-hand side of \Eqref{flowequd}.

For the purpose of this section, we now go back to our results for the
effective potential in the large-$N_\text{f}$ limit, in order to systematically
calculate corrections to this approximation. For this, we note that the bosonic
loop contribution is order $1/N_{\text{f}}$-suppressed compared to the fermionic
loop. This suggests a substitution of the large-$N_{\text{f}}$ solution into the
bosonic loop contribution of \Eqref{flowequd}, yielding an additional
contribution to the inhomogeneous part of the partial fixed point equation.

We also include the shift $\delta c_\epsilon=-1/(20\Nf)$ given in
\Eqref{eq:cepsNf} on the left-hand side of \Eqref{flowequd}. At fixed anomalous
dimensions and Yukawa coupling, this yields the particular solution $\delta
y=\delta c_\epsilon/(\eta_\phi-2)$ for $\eta_\phi\neq2$. The homogeneous
integration constant must be determined separately when imposing the same
matching condition on the leading and corrected solutions. This procedure can,
in principle, be iterated to any desired order.

In order to obtain analytic results, we perform this procedure in the asymptotic
regimes of small and large $\rho$ to order $1/N_{\text{f}}$. Note that the
fermionic loop satisfies the leading-order equation and as such cancels the
leading-order contribution of the linear, homogeneous part of the flow equation.
The resulting differential equation after cancelation therefore yields directly
the leading-order correction. 

In the large-$\rho$ regime, the correction consists of a
decaying bosonic contribution and a field-independent
solution associated with the shift in $c_\epsilon$:
\begin{equation}
\begin{aligned}
    \Delta y_{\text{l}}^{(1)} = {}&
    -\frac{
        (\eta_\phi-2)(2+\eta_\phi)^2
        \rho^{-\frac{4}{2+\eta_\phi}}
    }{64\pi^2(-6+\eta_\phi)}
    \\
    &\times \biggl[
        3(2+\eta_\phi)^{\frac{6+\eta_\phi}{2+\eta_\phi}}c
        -2^{2+\frac{4}{2+\eta_\phi}}\,
        d_\gamma (h^2)^{\frac{4}{2+\eta_\phi}}
        N_\text{f}v_4
    \\
    &\qquad \times
        (-6+\eta_\psi)
        \Gamma\left(-\frac{4}{2+\eta_\phi}\right)
        \Gamma\left(\frac{6+\eta_\phi}{2+\eta_\phi}\right)
    \biggr]^{-1}
    \\
    &+\frac{1}{20\Nf(2-\eta_\phi)}\,.
     \label{anasollrhocor}
\end{aligned}
\end{equation}
We observe that, for $0<\eta_\phi<2$, the scaling of the bosonic contribution
$\rho^{-4/(2+\eta_\phi)}$ causes a decay faster
than $1/\rho$ at large fields. The additional particular
solution associated with the correction to the mass flow
is independent of $\rho$. Both contributions are subleading
relative to the leading large-field behavior of $y(\rho)$.
Imposing the same matching condition on the leading and
corrected solutions also changes the coefficient of the
homogeneous contribution, while leaving its exponent
unchanged.

In the small-$\rho$ limit, the leading-order correction reads
\begin{equation}
\begin{aligned}
 \Delta y_{\text{s}}^{(1)}= {} & \frac{75 (-6 + \eta_\phi)^2 (-2 + \eta_\phi)^3 (2+\eta_\phi)^{
        \frac{2-\eta_\phi}{2+\eta_\phi}} c \,
  \rho^{- \frac{2 \eta_\phi}{2 + \eta_\phi}}}{64 \pi^2 (2 + \eta_\phi)^3} \\
    & \times \Bigl[ 3 (-6 + 5 \eta_\phi) + 10 N_\text{f} d_\gamma v_4 h^2 (-6 + \eta_\psi) \Bigr]^{-2} \\
    & + \frac{1}{20\Nf(2-\eta_\phi)} \, .
    \label{anasolsrhocor}
\end{aligned}
\end{equation}
For $\eta_\phi=2-\varepsilon$ with $0<\varepsilon\ll1$, the
leading-order correction diverges as $\rho^{-1+\varepsilon/4 +
\mathcal{O}(\varepsilon^2)}$ for small $\rho$. This is incompatible with the
finite boundary value $y(0)=\epsilon$ and also does not match the full numerical
results of Sect.~\ref{flowtoSSB}. We conclude that the small-$\rho$ expansion
taken together with the large-$\Nf$ expansion does not converge uniformly.
Nevertheless, the correction is integrable and yields a finite contribution to
the effective potential $u(\rho)$, which is compatible with the finite boundary
value $u(0)=0$. In other words, the large-$\Nf$ expansion introduces a
non-analytic behavior of $u(\rho)$ near the origin yielding diverging
derivatives at $\rho=0$. Thus, the correction scheme does not reliably describe
$y(\rho)$ arbitrarily close to $\rho=0$. Due to the rather small prefactor
$(\eta_\phi-2)^3\sim \varepsilon^3$, the correction remains numerically small
down to rather small values of $\rho$. This prefactor also guarantees that
non-analyticity disappears in the strict large-$\Nf$ limit, where $\eta_\phi\to2$.

Let us finally remark that the iterative scheme presented here for large and
small $\rho$ includes the large-$\epsilon$ correction to the mass flow together
with the bosonic contribution evaluated on the leading solution. A systematic
extension to even higher orders would also require the corresponding corrections
to $c_\epsilon$, the anomalous dimensions, the Yukawa coupling, and the field
dependence of the marginal direction.

\subsection{Alternative Approximation Schemes}

For the present system, alternative approximation schemes beyond the
large-$N_\text{f}$ limit can be based on the structure of the flow equation for
the effective potential \eqref{flowequd}. Solely relying on the argument given
below Eq.~\eqref{eq:partialFPdef2}, we can assume that the left-hand side of the
flow equation is constant, and thus the right-hand side must likewise be
constant. 

One potential realization of this balance is a direct cancelation between the
fermionic and bosonic loop contributions. For a growing power law $y(\rho)
\propto \rho^q$, the bosonic loop scales as $\rho^{-q-1}$. Cancelation against
the fermionic loop, which scales as $\rho^{-2}$, would therefore require $q=1$.
The dimensional scaling terms would then scale linearly with $\rho$, which is
incompatible with an (approximately) constant left-hand side of the fixed point
equation.

A second alternative scheme is obtained from a systematic analysis of the
large-$\rho$ regime of the effective potential. For large $\rho$, the fermionic
contribution becomes negligible due to its inverse scaling with the squared
invariant $\rho$. Consequently, only the linear, homogeneous part and the
bosonic contribution remain. For $0<\eta_\phi<2$, the leading large-field
solution generically scales as
$y(\rho)\propto\rho^{\frac{2-\eta_\phi}{2+\eta_\phi}}$. By inserting this
solution into the bosonic loop term, we find that this contribution scales as
$\rho^{\frac{-4}{2+\eta_\phi}}$ for large $\rho$. Because the bosonic part
vanishes in this limit, this constitutes a valid asymptotic solution. The
leading-order correction can be determined by analyzing the asymptotic behavior
of the bosonic contribution for large $\rho$ and solving the resulting partial
fixed-point equation.

For $\eta_\phi=2$, the large-field equation reduces to $4\rho y'(\rho)\simeq
c_\epsilon$. With $c_\epsilon=4/5$, this gives the logarithmic behavior in
Eq.~\eqref{anasolNflrho2}. Inserting this solution into the bosonic loop yields
a contribution that decays as $\frac{1}{\rho(\log\rho)^2}$, confirming the
consistency of the large-field approximation. Upon integrating with respect to
$\rho$, we find that the effective potential behaves as $\rho \log \rho$.
Remarkably, this reproduces the exact same behavior found in the mean-field
approximation \cite{Gies:2024dly}. This result is entirely self-consistent, as we initially assumed
an anomalous dimension of $\eta_\phi = 2$ and a left-hand side value of $4/5$,
which go hand in hand with the results of the large-$N_\text{f}$ limit. 

As a third alternative scheme, one could also start from the
assumption that the denominator of the bosonic loop $(1+y + 2 \rho y')^2$ is
proportional to $\rho^n$ with $n \in \mathbb{R}$, thereby necessitating a
constant contribution of $y\to -1$. This can be implemented  explicitly by
introducing the shifted variable
\begin{equation}
    z = y + 1\,.
\end{equation}
In terms of $z$, the flow equation reduces to
\begin{equation}
\begin{aligned}
    c_\epsilon = \, & (-2+\eta_\phi) (z-1) + (2+\eta_\phi) \rho z' \\
    & - 2v_4 \Bigg[\frac{1}{2}\left(1-\frac{\eta_\phi}{6} \right)\frac{3 z'+2 \rho z''}{(z+2 \rho z')^2} \\
    & - N_\text{f} d_\gamma \left(1-\frac{\eta_\psi}{6} \right)\frac{2 h^2}{(1+2 \rho h^2)^2}\Bigg]\,.
\end{aligned}
\end{equation}
In order to ensure a constant contribution from the right-hand side, the linear,
homogeneous and the fermionic loop contributions must not grow with $\rho$. By
construction, if one of these contributions were to grow, the other would remain
bounded, making it impossible to cancel each other.

Thus, a balanced, constant solution on both sides of the differential equation
can only be established by considering an additional term of the form
$\frac{c}{\rho}$. There, the coefficient $c$ is uniquely fixed to match the
value of $c_\epsilon$ on the left-hand side. We therefore make the ansatz
\begin{equation}
    x = z - \frac{c}{\rho}\,,
\end{equation}
where $x(\rho)$ encapsulates subleading contributions. Keeping only
the leading terms, $z(\rho)=c/\rho$, we obtain
\begin{equation*}
    \frac{3 z'+2 \rho z''}{(z+2 \rho z')^2} = \frac{1}{c}\,.
\end{equation*}
Matching the constant terms in the large-$\rho$ limit therefore gives
\begin{equation}
    c_\epsilon = 2-\eta_\phi
    - \frac{v_4}{c}\left(1-\frac{\eta_\phi}{6}\right)\,.
\end{equation}
For $c_\epsilon=4/5$, this fixes
\begin{equation}
    c = \frac{v_4(1-\eta_\phi/6)}{6/5-\eta_\phi}\,,
\end{equation}
which is negative for $6/5<\eta_\phi<6$. The expansion can be continued
by canceling the subleading contributions order by order.

The leading terms already show that the potential is concave
at large fields as a function of the dimensionless renormalized
scalar field $\varphi=\sqrt{Z_\phi}\,\phi/k$.
Indeed, with $\rho=\varphi^2/2$, its curvature satisfies
\begin{equation}
    \frac{d^2}{d\varphi^2}u(\varphi^2/2)
    = y(\rho)+2\rho y'(\rho)
    \longrightarrow -1
    \qquad \text{as }\rho\to\infty.
\end{equation}
Moreover, since $y(\rho)\to-1$ as $\rho\to\infty$, the potential behaves as
$u(\rho)\sim-\rho$ at large fields and is therefore unbounded from below. We
therefore discard this solution provided by the third scheme.

\section{Numerical solution of the functional flow}
\label{sec:Numerical}

Let us now turn to the numerical solution of the functional flow within the
truncated theory space, but without relying on the large-$N_\text{f}$
approximation or field expansions. For the remainder of this section, we continue working
exclusively in $d=4$ dimensions, with $d_\gamma = 32$ and $v_4 =
\frac{1}{32 \pi^2}$.

\subsection{Partial Fixed-Point Solution}

As a first step, we determine the partial fixed-point solution of the flow
equations for the effective potential, using the analytical results of the
previous section as a guide. E.g., from the preceding section and in agreement
with a polynomial expansion, we expect the minimum of the partial fixed-point
potential to be located in the symmetric regime, i.e., $\epsilon > 0$ and $\kappa = 0$. 

For any given value of $\epsilon>0$, both anomalous dimensions and the
fixed-point value of the Yukawa coupling can be determined analytically by solving
the coupled system of equations arising from the Yukawa flow \eqref{eq:flowhq} and the explicit
expressions \eqref{eq:etapsi}, \eqref{eq:etaphi} for the anomalous dimensions.

\begin{figure}[t]
        \centering
        \includegraphics[width=0.98\linewidth,trim={0.7cm 0.25cm 0.5cm 0.2cm}]{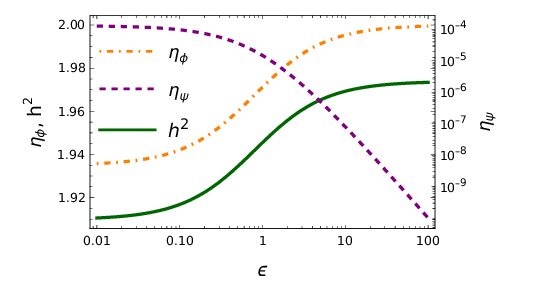}
        \caption{Dependence of the anomalous dimensions $\eta_\phi$, $\eta_\psi$
        and the Yukawa coupling $h^2_*$ at the partial fixed point on the
        renormalized mass parameter $\epsilon$ for $N_\text{f} = 1$.}
        \label{fig:epsdep}
\end{figure}

Notably, the specific value of $\epsilon$ introduces only a small dependence on
the anomalous dimensions and the Yukawa coupling. A larger value of the
renormalized mass parameter $\epsilon$ shifts the anomalous dimensions closer to
the large-$N_\text{f}$ limit, where $\eta_\phi = 2$ and $\eta_\psi = 0$. For all
finite values of $\epsilon$, $\eta_\phi$ remains slightly below $2$ while $\eta_\psi$
stays marginally above $0$. Furthermore, the Yukawa coupling increases for
larger values of the renormalized mass parameter $\epsilon$. This behavior is
depicted in Fig.~\ref{fig:epsdep}, which displays the dependence of
$\eta_\phi$, $\eta_\psi$, and $h^2_*$ on $\epsilon$ for $N_\text{f} = 1$.

\begin{figure}[t]
     \centering
     \includegraphics[width=0.98\linewidth,trim={0.7cm 0.25cm 0.5cm 0.2cm}]{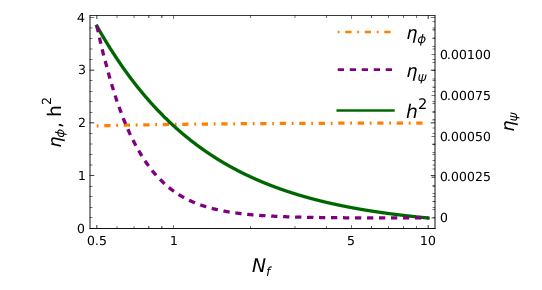}
     \caption{Dependence of the anomalous dimensions $\eta_\phi$, $\eta_\psi$ and the Yukawa coupling $h^2_*$ at the partial fixed point on $N_\text{f}$ for $\epsilon = 10$.}
     \label{fig:Nfdep}
\end{figure}

These partial fixed-point properties also behave as expected as a function of
$N_\text{f}$ at a fixed renormalized mass parameter $\epsilon$: the anomalous
dimensions exhibit only a weak dependence on $N_\text{f}\geq 1$, with higher
values of $N_\text{f}$ systematically approaching the asymptotic limits
$\eta_\phi = 2$ and $\eta_\psi = 0$. The Yukawa coupling $h^2_*$ displays a
pronounced inverse scaling with $N_\text{f}$ and vanishes entirely in the
large-$N_\text{f}$ limit; however, for small flavor numbers $N_\text{f} \sim
O(1)$, the Yukawa coupling is also of order $O(1)$ signifying a comparatively
strong interaction. Overall, the behavior of the Yukawa coupling  at its partial
fixed point is in excellent agreement with the large-$N_\text{f}$ expectation
$h^2_* \sim 1/N_\text{f}$ and is illustrated in Fig.~\ref{fig:Nfdep}, which
depicts the dependence of $\eta_\phi$, $\eta_\psi$, and $h^2_*$ on $N_\text{f}$
for a fixed renormalized mass parameter $\epsilon = 10$.

The proximity of the anomalous dimensions to their large-$N_\text{f}$ limits for
a wide range of both $N_\text{f}$ and $\epsilon$ values arises from the
dominance of the fermionic loop contributions. This agreement justifies the
large-$N_\text{f}$ approximation in this regime -- even for small values of $N_\text{f}$.

\begin{figure}[t]
    \centering
    \includegraphics[width=\linewidth]{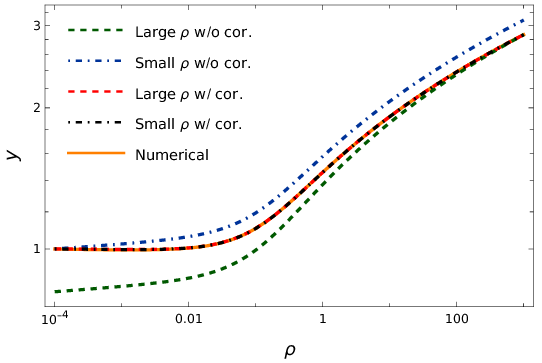}
    \caption{The derivative of the effective potential $y(\rho)$ as a function
    of the field invariant $\rho$ at the partial fixed point for $N_{\text{f}} =
    1$ and renormalized mass parameter $\epsilon = 1$. The full numerical
    solution is compared to the analytical approximations obtained in the
    large-$N_{\text{f}}$ limit, as well as approximations incorporating
    leading-order $1/\Nf$ corrections $\Delta
y_{\text{l}}^{(1)}$ from the large-$\rho$ expansion, cf. \Eqref{anasollrhocor},
  using matching conditions at small and large $\rho$.}
    \label{fig:effpot}
\end{figure}

For the subsequent numerical analysis, we choose the renormalized mass parameter
as $\epsilon = 1$ by way of example. For $N_{\text{f}} = 1$, this choice yields
the anomalous dimensions $\eta_\phi \approx 1.971$ and $\eta_\psi \approx 1.479
\times 10^{-5}$, alongside the fixed-point Yukawa coupling $h^2_* \approx
1.946$. We determine the numerical partial fixed-point solution from
\Eqref{flowequd} using the constant approximation $\partial_t
y\simeq4/5-1/(20\Nf)$ from the preceding section. For $\Nf=1$, this gives
$c_\epsilon=0.75$. This approximation neglects further $\rho$ and $\epsilon$
dependencies as well as higher-order $N_\text{f}$ corrections. This
approximation turns out to be quite accurate when comparing to the full
numerical flow, as done in Section \ref{flowtoSSB}. The resulting numerical
solution to the partial fixed-point equation is displayed in
Fig.~\ref{fig:effpot}, where it is compared to the corresponding analytical
approximations obtained in the large-$N_{\text{f}}$ limit.

For the analytical solution, the bosonic sector can be omitted because it is
 highly suppressed for $\eta_\phi \approx 2$. The large-field expression in
Eq.~\eqref{anasollrhocor} contains an explicit factor
$(\eta_\phi-2)$ in the numerator. Since these
suppressions persist for higher-order corrections, the large-$N_{\text{f}}$
limit provides a robust approximation even for $\Nf=1$.

To determine the integration constant $c$ of the analytical solution, we match
the analytical and numerical solutions for $y(\rho)$ in the deep asymptotic
large-$\rho$ regime at $\rho = 1000$. This choice is motivated by the observation that both
fermionic and bosonic loops become negligible in this regime, leaving only the
linear, homogeneous part of the differential equation. Because our analytical
approximation explicitly accounts for this asymptotic behavior, imposing the
matching condition at large $\rho$ is both physically and mathematically
justified.

Alternatively, we can match the values of
$y(\rho)=u'(\rho)$ also in the small-$\rho$ regime at $\rho=10^{-4}$. The integration
constant $c$ is determined separately for each analytical
approximation, including the corrected ones.

As illustrated in Fig.~\ref{fig:effpot}, the analytical
large-$N_{\text{f}}$ approximations already describe the numerical
solution quite well. Matching in the large-$\rho$ regime gives
good qualitative agreement for all $\rho$ and accurate quantitative
agreement at large fields. At small $\rho$, the leading
approximation underestimates $y(\rho)$, mainly because of
the omission of the correction $-1/(20\Nf)$ on the left-hand side
of the flow equation. Matching at small $\rho$ also gives
good qualitative agreement for all $\rho$, but tends to overestimate
$y(\rho)$ away from the matching point. Nevertheless, both
matching prescriptions demonstrate that the large-$N_{\text{f}}$
approximation captures the main features of the effective
potential.

Notably, incorporating explicit analytical $1/\Nf$ corrections $\Delta
y_{\text{l}}^{(1)}$ from the large-$\rho$ expansion, cf. \Eqref{anasollrhocor},
yields an important quantitative improvement. This improvement appears to arise
mainly from including the correction to the mass flow, while the bosonic
contribution is significantly smaller. The corrected approximation is
illustrated in Fig.~\ref{fig:effpot}, where the large-$N_{\text{f}}$
approximation is supplemented with the large-field corrections derived in
Sect.~\ref{slfieldexp}. To ensure a consistent comparison, the same matching
conditions are imposed on the leading and corrected approximations, with the
integration constant determined separately in each case. Incidentally, $\Delta
y_{\text{l}}^{(1)}$ captures the dominant leading $1/\Nf$ correction only in the
very large-$\rho$ regime. At intermediate values of $\rho$, the leading-order
correction is dominated by non-asymptotic, i.e., power-law type contributions
which remain quantitatively below a one percent level even for $\Nf=1$.

\begin{figure}[t]
    \centering
    \includegraphics[width=0.48\textwidth]{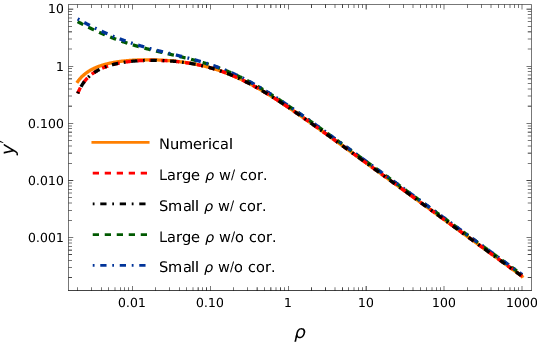}
    \caption{The second derivative of the effective potential $y'(\rho)$ as a
    function of the field invariant $\rho$ at the partial fixed point for
    $N_{\text{f}} = 1$ and renormalized mass parameter $\epsilon = 1$. The full numerical
    solution is compared to the analytical approximations obtained in the
    large-$N_{\text{f}}$ limit, as well as approximations incorporating
    leading-order $1/\Nf$ corrections $\Delta
y'{}_{\text{l}}^{(1)}$ from the large-$\rho$ expansion, cf. \Eqref{anasollrhocor}, 
  using matching conditions at small and large $\rho$.}
    \label{fig:effpotd}
\end{figure}

Furthermore, we evaluate the second derivative of the effective potential,
$y'(\rho)$, as shown in Fig.~\ref{fig:effpotd}. This observable is particularly
insightful because the approximation $c_\epsilon = 4/5 - 1/(20 \Nf)$ drops out of the flow
equation for $y'(\rho)$.

Figure~\ref{fig:effpotd} shows that the inclusion of $1/\Nf$ corrections
significantly improves the quantitative agreement for small $\rho$, although
some deviations remain. Specifically, the quantitative agreement of the
large-$\rho$ expansion with the numerical solution even far outside its expected
validity regime is remarkable.
The improvement from the $1/\Nf$ corrections may again be largely
associated with the correction to the mass flow, which affects
$y'(\rho)$ through the readjustment of the matching constant.
At larger $\rho$, all curves agree to high accuracy, indicating that
the corrections have only a small effect in this regime.
For $\rho\gtrsim1$, the curves become approximately straight in the
double-logarithmic plot, consistent with the expected large-field
scaling $y'(\rho)\propto\rho^{-2\eta_\phi/(2+\eta_\phi)}$.

\begin{figure}[t]
    \centering
    \includegraphics[width=\linewidth]{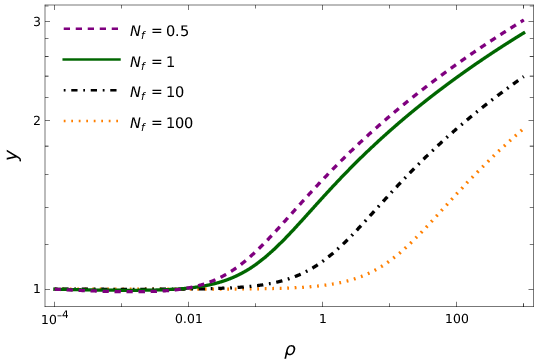}
    \caption{The derivative of the effective potential $y(\rho)$ as a function
    of the invariant $\rho$ for a fixed mass parameter $\epsilon = 1$ and
    various values of $N_{\text{f}}$. The delayed point of increase
    towards larger $\rho$ is in line with the $\Nf$ scaling deduced in \Eqref{Nfsc}.}
    \label{fig:effpotNf}
\end{figure}

\begin{figure}[t]
    \centering
    \includegraphics[width=\linewidth]{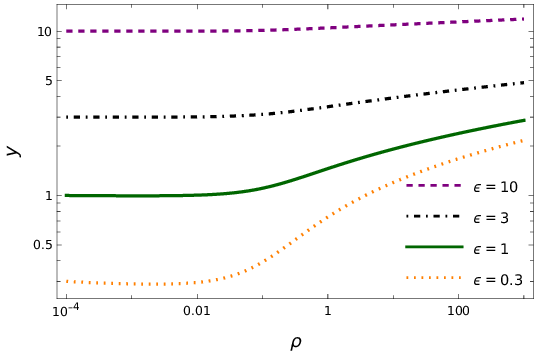}
    \caption{The derivative of the effective potential $y(\rho)$ as a function
    of the invariant $\rho$ for $N_{\text{f}} = 1$ and various choices of the
    renormalized mass parameter $\epsilon$.}
    \label{fig:effpoteps}
\end{figure}

Finally, we compare the numerical solutions across various values of
$N_\text{f}$ and $\epsilon$, adjusting the anomalous dimensions and the partial
fixed-point Yukawa coupling accordingly. These comparative results are presented
in Fig.~\ref{fig:effpotNf} and Fig.~\ref{fig:effpoteps}, respectively.

The qualitative behavior remains stable across different values of
$N_{\text{f}}$ and $\epsilon$. Notably, a larger value of $N_{\text{f}}$ delays
the onset of the strong growth in the derivative of the effective potential.
Nevertheless, the functional dependence of $y(\rho)$ on large values of the
invariant $\rho$ is mostly independent of the specific choices of $N_{\text{f}}$
and $\epsilon$ within the considered parameter range. Furthermore, because
$y(\rho)$ remains strictly monotonically increasing for all values of $\rho$ in
all cases, we conclude that the effective potential satisfies the conditions
implied by the expected convexity. It also  confirms the absence of a second
minimum, justifying the assumptions $\kappa = 0$ and $u'(\kappa) = \epsilon$
throughout our calculations.

\subsection{Flow to SSB regime}
\label{flowtoSSB}

We now investigate the full flow, starting in the symmetric regime and
transitioning to the SSB regime, including the infrared limit $k \rightarrow
0$. We numerically solve the full, coupled system of flow equations for the
effective potential and the Yukawa coupling without further approximations. The
numerical implementation is written in Julia, partially inspired by existing
implementations from \cite{Ihssen:2023qaq}, building on various numerical methods for FRG equations
developed in the past \cite{Grossi:2019urj,
Koenigstein:2021syz,Ihssen:2022xkr,Zorbach:2024rre,Zorbach:2024zjx}. The code heavily relies on
\texttt{DifferentialEquations.jl} \cite{Rackauckas2017} and \texttt{Symbolics.jl} \cite{Gowda2021}. Additional
interpolation and root-finding functionality is provided by
\texttt{DataInterpolations.jl} \cite{Bhagavan2024} and \texttt{Roots.jl} \cite{Verzani2020}, respectively. The flow is
integrated using the \texttt{KenCarp58} algorithm \cite{KENNEDY2019183}, which
is well-suited for potentially stiff systems of renormalization group equations,
as shown, for example, in \cite{Ihssen:2023qaq}.

Before comparing our results for different $N_\text{f}$, we first analyze the
flow of $N_\text{f}=1$ in more detail. We use
the ultraviolet cutoff to define the  scale $\Lambda = 1.0$. The initial conditions at the
ultraviolet scale are chosen to be $\epsilon_\Lambda = 10$, $\lambda_\Lambda =
1$, and $h^2_\Lambda = 1$ for the renormalized mass parameter, the quartic
coupling, and the squared Yukawa coupling, respectively. Additionally, we use
$Z_\phi = 1.0$ and $Z_\psi = 1.0 $ for the wave function renormalization
factors. The initial effective potential is simply of $\phi^4$ type, satisfying
\begin{equation}
    u_\Lambda'(\rho) = \epsilon_\Lambda + \lambda_\Lambda \rho,
\end{equation}
where the coupling $\lambda_\Lambda$ must be positive to ensure that the
effective potential is bounded from below.

To evaluate the field dependencies of the effective potential for different
renormalization scales, we discretize the field space invariant $\rho$ on an
equally spaced grid consisting of $n = 3000$ points. The grid is defined on the
interval $\rho \in [0, 1]$. Rather than evolving the potential $u(\rho)$ itself,
we evolve its first derivative, $y(\rho)=u'(\rho)$, at every grid point.

The flow equations are integrated using an adaptive step-size algorithm as
implemented in \texttt{DifferentialEquations.jl}. The internal step size in the
renormalization time $t$ is dynamically adjusted according to the estimated
local integration error and the prescribed tolerances $\texttt{atol}=10^{-13}$
and $\texttt{rtol}=10^{-11}$. 
Consequently, the numerical step size is not fixed throughout the evolution and
can be reduced automatically in regions where the flow exhibits rapid scale
dependence, particularly in the vicinity of the transition to the broken regime. 

To investigate the symmetric regime, we employ dimensionless quantities. During
every evaluation of the right-hand side, we construct a cubic spline interpolant
from the discrete values of $y(\rho)$. This is used to evaluate the first and
second field derivatives required in the flow equations and in the anomalous
dimensions. In particular, at the origin, the quantities $y'(0)$ and $y''(0)$
are obtained from the spline interpolation. They are then inserted into the
expressions for $\eta_{\phi}$ and $\eta_{\psi}$. Those expressions for the
anomalous dimensions are deduced from solving the coupled system of equations
given by \Eqref{eq:etaphi} and \Eqref{eq:etapsi}. The anomalous dimensions are
therefore evaluated self-consistently from the instantaneous values of the
effective potential, Yukawa coupling, and the derivatives of the potential. 

A critical feature of our numerical setup is the dynamic detection of the
transition from the symmetric to the broken regime. We implement a discrete
callback that monitors the mass parameter at the origin. Once this mass
parameter drops to zero or below ($y(\rho = 0) \le 0$), indicating the onset of
spontaneous symmetry breaking, the system transitions to the broken regime. This
transition occurs at $t_\text{SSB} \approx -26.21$, where we have $h^2 \approx
1.924$ and $\lambda \approx 1.429$.  In the broken regime, we use dimensionful
quantities which we measure in units of the  dimensionful scale $k_{\text{SSB}}=
\Lambda e^{t_{\text{SSB}}}$ as a reference scale for all dimensionful quantities
in the following. We ultimately follow the flow until $t = -30.8$. A Boolean flag
switches the right-hand side of the ODE system from the dimensionless flow
equations of the symmetric regime to the dimensionful ones used in the broken
regime. Hence, the two regimes are treated using different parametrizations
while remaining part of a single numerical evolution. 

Within the broken regime, the flow equations are formulated entirely in terms of
dimensionful quantities. Because the minimum of the effective potential moves
away from the origin once the system enters the SSB regime, the flowing minimum
$\kappa$, defined via $y(\kappa) = 0$, needs to be determined dynamically.
Therefore, we locate $\kappa$ at every integration step. We identify the pair of
neighboring grid points at which $y(\rho)$ changes sign and obtain $\kappa$ from
a linear interpolation between them. The curvature quantities $y'(\kappa)$ and
$y''(\kappa)$ entering the anomalous dimensions, Eqs.~\eqref{eq:etaphi} and
\eqref{eq:etapsi}, are subsequently evaluated at this dynamically determined
minimum using the cubic spline interpolant.

The spline-based evaluation of the field derivatives used to solve the flow
equations in the symmetric regime becomes numerically unstable once a
non-trivial minimum develops, as $y(\rho)$ forms a pronounced kink in its
vicinity. Following the strategy outlined in \cite{Ihssen:2023qaq}, once the
system has entered the broken regime, we discretize the right-hand side of the
potential flow directly on the grid via finite differences between neighboring
grid points, using one-sided differences at the boundaries $\rho_\text{min}$ and
$\rho_\text{max}$. With this discretization, we obtain a set of coupled ordinary
differential equations.

Following the first-order finite difference scheme described in
\cite{Grossi_2021}, we define the upwind derivative as
\begin{equation}
    \mathbf{D}_\text{u} f_i = \frac{f_{i+1}-f_i}{\Bar{\rho}_{i+1}-\Bar{\rho}_i}
\end{equation}
and the downwind derivative as
\begin{equation}
    \mathbf{D}_\text{d} f_i = \frac{f_{i}-f_{i-1}}{\Bar{\rho}_{i}-\Bar{\rho}_{i-1}}.
\end{equation}
Using these definitions, the flow can be expressed as
\begin{equation}
\begin{aligned}
    \pat Y(\bar{\rho})
    ={}& 2 v_d k^{d+2} \mathbf{D}_{\mathrm{u}} \Bigg[
    \frac{2}{d}\left(1-\frac{\eta_\phi}{d+2}\right)
    \\
    &\quad \times
    \frac{1}{
        k^2+\frac{Y}{Z_\phi}
        +2\frac{\bar{\rho}}{Z_\phi}\mathbf{D}_{\mathrm{d}}Y
    }
    \\
    &\quad
    -N_{\mathrm{f}}d_\gamma\frac{4}{d}
    \left(1-\frac{\eta_\psi}{d+2}\right)
    \frac{k^2}{
        k^4+2\bar{\rho}\frac{\bar{h}^2}{Z_\psi^2}
    }
    \Bigg].
\end{aligned}
\end{equation}
This expression holds in general spacetime dimension $d$, with $\bar{\rho}$
denoting the dimensionful field invariant, $\bar{h}^2$ the dimensionful
squared Yukawa coupling, and $Y(\bar{\rho})=\partial U/\partial\bar{\rho}$
the dimensionful derivative of the effective potential. In the following,
we consider $d=4$.

For the boundary conditions at the grid cutoff $\Bar{\rho}_\text{max}$, we
assume that the upwind and downwind derivatives are interchangeable and thus
replace the upwind derivative with a downwind derivative. This is justified
because, for very large fields, loop contributions are heavily suppressed.
Furthermore, we impose that $2 \Bar{\rho}\, \mathbf{D}_\text{d} Y$ vanishes at
$\Bar{\rho} = 0$, assuming the effective potential is regular here.

\begin{figure}[htbp]
    \centering
    \includegraphics[width=\linewidth]{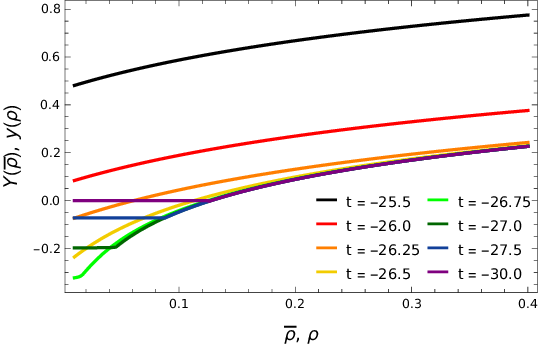}
    \caption{Flow of the derivative of the dimensionful effective
    potential $Y(\Bar{\rho})$ in the broken regime and dimensionless effective
    potential $y(\rho)$ in the symmetric regime in terms of the field invariants
    $\Bar{\rho}$ and $\rho$, respectively, at different values of the
    renormalization group time $t$ for $N_\text{f}=1$, starting from the
    initial conditions $\epsilon_\Lambda=10$, $h_\Lambda^2=1$,
    $\lambda_\Lambda=1$. The upper two curves correspond to the
    symmetric regime, where the potential has approached the partial fixed point
    and the residual flow is driven by the mass parameter, corresponding to a
    constant shift of the potential derivative. The lower curves exhibit the
    flow in the broken regime, where the potential develops a non-trivial
    minimum and convexity is restored as $k \rightarrow 0$ ($t\to - \infty$).}
    \label{fig:effpot_flow}
\end{figure}

Figure~\ref{fig:effpot_flow} illustrates the RG scale dependence of the
dimensionless effective potential $y(\rho)$ in the symmetric regime together
with that of the dimensionful effective potential $Y(\Bar{\rho})$ in the broken
regime for several representative values of $k$ near the transition to the
broken regime at $t_\text{SSB}$. For $t\gtrsim t_\text{SSB}$, the potential has
approached the partial fixed point and the residual flow is driven by the mass
parameter, corresponding to a $\rho$-independent constant shift of the potential
derivative. Below the transition, the potential develops a non-trivial minimum
and convexity is restored as $k \rightarrow 0$.

\begin{figure}[htbp]
    \centering
    \includegraphics[width=\linewidth]{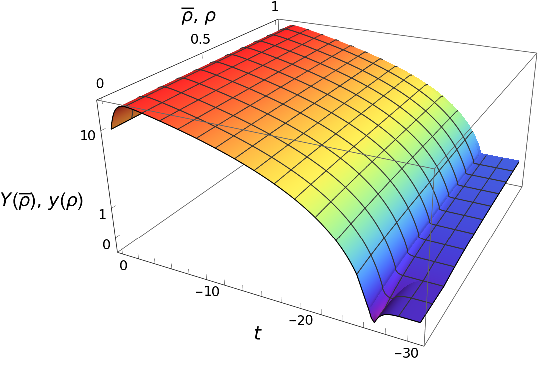}
    \caption{Three-dimensional plot of the derivative of the dimensionful
    effective potential $Y(\Bar{\rho})$ in the broken regime and dimensionless
    effective potential $y(\rho)$ in the symmetric regime in terms of the field
    invariants $\Bar{\rho}$ and $\rho$, respectively, and the renormalization
    group time $t$ for $N_\text{f}=1$; initial conditions as in
    Fig.~\ref{fig:effpot_flow}. In the initial part of the flow $k\lesssim
    \Lambda$, ($t\lesssim 0$), the potential starts in the symmetric regime and
    quickly approaches the partial fixed point. Over a wide range of scales, the
    potential evolves only through the logarithmic running of the mass
    parameter, until the transition to the broken regime occurs at naturally
    small scales $k\ll \Lambda$ ($t\simeq -26$). In the broken regime, the
    potential develops a non-trivial minimum.} 
    \label{fig:effpot_flow_3D}
\end{figure}

Figure~\ref{fig:effpot_flow_3D} provides a three-dimensional representation of
these potential derivatives as functions of their respective field invariants
($\Bar{\rho}$ and $\rho$) over a large range of the RG time $t$. During the
initial part of the flow $k\lesssim \Lambda$ ($t\lesssim 0$), where the
potential starts from $\phi^4$ form with $y(\rho)$ being strictly linear, the
potential quickly approaches the partial fixed point. Over a wide range of
scales, the potential evolves only through the logarithmic running of the mass
parameter, until the transition to the broken regime occurs at naturally small
scales $k\ll \Lambda$ ($t\simeq -26$). In the broken regime, the potential
develops a non-trivial minimum of the order of the symmetry-breaking transition scale.


\begin{figure}[htbp]
    \centering
    \includegraphics[width=\linewidth]{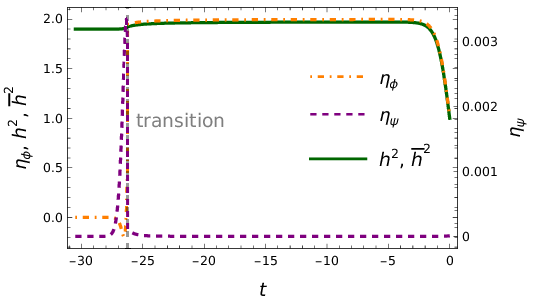}
    \caption{RG flow of the anomalous dimensions $\eta_{\phi}$ and $\eta_{\psi}$
    and the squared Yukawa coupling $h_k^2$ from the symmetric to the broken
    regime for $N_\text{f}=1$; initial conditions as in
    Fig.~\ref{fig:effpot_flow}. Before the transition, we show the dimensionless squared Yukawa
coupling, and after the transition, the dimensionful squared
Yukawa coupling in units of $k_{\text{SSB}}^2$.}
    \label{fig:yuk_anomdim}
\end{figure}

Figure~\ref{fig:yuk_anomdim} depicts the flow of the Yukawa coupling $h^2$ and
the anomalous dimensions $\eta_\phi$ and $\eta_\psi$ as functions of $t$,
tracking the transition from the symmetric to the broken regime. For the Yukawa
coupling, we employ dimensionless quantities in the symmetric regime, where it
approaches its fixed-point value, and dimensionful quantities in the broken
regime, where it freezes out. As $k$ decreases, the Yukawa coupling
drops slightly, the dimensionful freeze-out value is slightly
smaller than the partial fixed point value (in units of $k_{\text{SSB}}^2$). The
anomalous dimensions initially flow towards their partial fixed-point values
in the symmetric regime, run fast near $k_{\text{SSB}}$, and approach zero for
$k \rightarrow 0$. Notably, at the transition point, the anomalous dimension
$\eta_\phi$ exhibits a sharp decline, which is driven by the minimum of the
potential shifting away from the origin.

\begin{figure}[htbp]
    \centering
    \includegraphics[width=\linewidth]{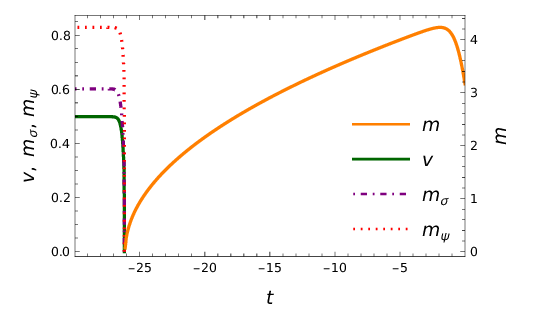}
    \caption{Flow of the dimensionful vacuum expectation value $v$, the
    dimensionful scalar mass $m_\sigma$, and the dimensionful fermionic mass gap
    $m_\psi$ in the broken regime in units of $k_{\text{SSB}}$, as well
    as the dimensionless mass parameter $m$ in the symmetric regime for
    $N_\text{f}=1$; initial conditions as in
    Fig.~\ref{fig:effpot_flow}.}
    \label{fig:vev_flow}
\end{figure}

\begin{figure}[htbp]
    \centering
    \includegraphics[width=\linewidth]{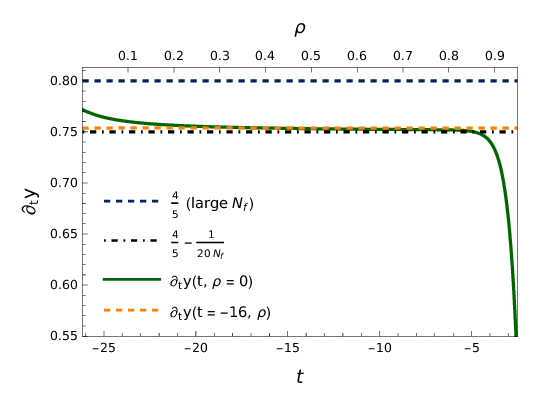}
    \caption{RG-time dependence of $\partial_t y$ at $\rho=0$
    and field dependence of $\partial_t y(\rho)$ at $t=-16$ for
    $N_\text{f}=1$, compared with the analytic large-$N_\text{f}$
    prediction of $4/5$ and the corrected prediction $4/5-1/(20N_\text{f})$.
    The initial conditions are the same as in Fig.~\ref{fig:effpot_flow}.}
    \label{fig:c_epsilon_flow}
\end{figure}

Figure~\ref{fig:vev_flow} displays the flow of the infrared observables: the
vacuum expectation value $v$, the scalar mass $m_\sigma$, and the fermionic mass
gap $m_\psi$ in the broken regime.
The values of the coupling $\lambda$ entering the scalar mass were
obtained using finite differences and agree with those obtained by
differentiating (cubic) spline interpolants.
Once the flow is in the broken regime, these
quantities grow continuously from zero. For comparison, the dimensionless mass
parameter $m$ is shown in the symmetric regime, where its square $m^2=\epsilon$ runs
logarithmically;
this is in excellent agreement with our analytic approximation of $\partial_t
\epsilon = 4/5 - 1/(20 \Nf)$ at the partial fixed point. 
More quantitatively, we can see this behavior more clearly in
Figure~\ref{fig:c_epsilon_flow}, which shows $\partial_t y$ for $N_\text{f}=1$
as a function of $t$ at $\rho=0$ (solid green line) where the flow is evaluated numerically using
\begin{equation}
    \partial_t y\left(\frac{t_i+t_{i+1}}{2},\rho=0\right)
    \simeq
    \frac{\epsilon(t_{i+1})-\epsilon(t_i)}{t_{i+1}-t_i}\,.
\end{equation}
Within the partial fixed-point regime $t\lesssim -5$, $\partial_t y(t,\rho=0)$
increases slightly from approximately $0.75$ at large $\epsilon$
to $0.77$ as $\epsilon\to0$. The large-$N_\text{f}$ prediction of
$4/5$ thus overestimates the numerical result by about $4 - 7\%$.
The large-$\epsilon$ value agrees well with the perturbative result
of $c_\epsilon=0.75$ for $N_\text{f}=1$ obtained in
Ref.~\cite{Gies:2025pqv} using a quartic scalar potential.

In the same Fig.~\ref{fig:c_epsilon_flow}, we also show $\partial_t y$ as a function of
$\rho$ (dashed orange line) at $t=-16$, an RG time well within the partial fixed-point regime.
This is obtained from the right-hand side of
Eq.~\eqref{flowequd} using the interpolated $y(\rho)$, the anomalous
dimensions $\eta_\phi$ and $\eta_\psi$, and the Yukawa coupling $h^2$,
all extracted from the full numerical solution at this fixed RG time. 

At fixed RG time, $\partial_t y(t=-16,\rho)$ remains approximately constant,
varying between $0.75369$ and $0.75380$ throughout the displayed
field range. The large-$N_\text{f}$ prediction $4/5$ overestimates
the numerical result by about $6.1\%$, whereas the corrected
prediction $4/5-1/(20N_\text{f})=0.75$ underestimates it by only
$0.5\%$.

We interpret the weak field dependence of $\partial_t y(t=-16,\rho)$,
with a relative variation of less than $0.02\%$, as a justification
for our functional definition of the partial fixed point in
\Eqref{eq:partialFPdef2} (and the discussion below this equation).
This supports using a constant approximation for $\partial_t y$
in our analytic partial fixed-point analysis, with the leading
$1/N_\text{f}$ correction substantially improving the quantitative
agreement relative to the large-$N_\text{f}$ value $4/5$.

For $N_\text{f}=1$, the flow reaches $k/k_{\mathrm{SSB}} \simeq 10^{-2}$
at $t=-30.8$. Reading off the mass ratios at this scale and extrapolating
the infrared flow to $k=0$ give the same estimates. We obtain
\begin{equation}
   \Nf=1:\quad \frac{m_\sigma}{v} \approx 1.20, \quad \frac{m_\psi}{v} \approx 1.65, \quad \frac{m_\sigma}{m_\psi} \approx 0.72.
\end{equation}
We emphasize that these ratios are universal and largely independent of the
initial conditions, provided that the flow has spent a sufficiently long RG time
near the partial fixed point. As quantified in \cite{Gies:2025pqv}, this can be
achieved for a wide range of generic initial conditions without the need for any
fine-tuning.

\begin{figure}[htbp]
    \centering
    \includegraphics[width=\linewidth]{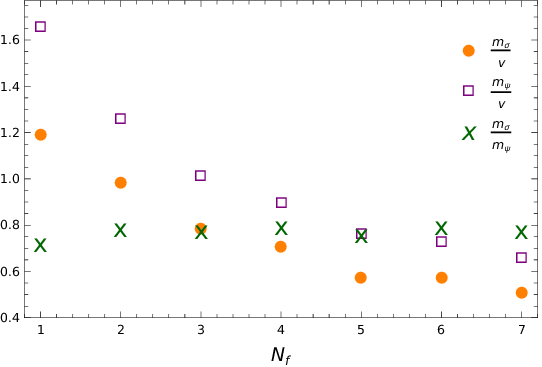}
    \caption{Infrared mass ratios $m_\sigma/v$, $m_\psi/v$, and $m_\sigma/m_\psi$ as functions of $N_\text{f}$. The values at the final integration scale
agree with those obtained by extrapolating the flow to $k=0$.}
    \label{fig:massratios}
\end{figure}

Repeating the same numerical calculation for different values of $N_\text{f}$,
we observe, as illustrated in Figure~\ref{fig:massratios}, that
$\frac{m_\sigma}{v}$ and $\frac{m_\psi}{v}$ decrease for higher $N_\text{f}$. As
we can deduce from Eq.~\eqref{Nfsc}, both scale with the inverse of
$\sqrt{N_\text{f}}$, since $\lambda$ and the Yukawa coupling $h^2$ scale with
the inverse of $N_\text{f}$, while $\kappa$ scales directly with $N_\text{f}$.
As a consequence, the ratio $\frac{m_\sigma}{m_\psi}$, as seen in
Figure~\ref{fig:massratios}, is rather insensitive to $N_\text{f}$, because the
scalings of the Yukawa coupling and the quartic coupling with $N_\text{f}$
cancel out. E.g. for $\Nf=7$, we find
\begin{equation}
\Nf=7:\quad \frac{m_\sigma}{v} \approx 0.51, \quad \frac{m_\psi}{v} \approx 0.66, 
   \quad \frac{m_\sigma}{m_\psi} \approx 0.78.
\end{equation}
For $N_\text{f}=3,\ldots,7$, the quantities
$\sqrt{N_\text{f}}\,m_\sigma/v$, $\sqrt{N_\text{f}}\,m_\psi/v$,
and $m_\sigma/m_\psi$ deviate from their respective mean values
by at most $6.0\%$, $3.1\%$, and $2.9\%$, respectively.
This scatter may reflect numerical uncertainty as well as
finite-$N_\text{f}$ corrections. The latter arise from bosonic fluctuations
and subleading contributions to the Yukawa flow and anomalous
dimensions. Calculations with more grid points agree with the displayed results,
supporting their stability under a refinement of the grid.

Comparing the mass ratio $\frac{m_\sigma}{m_\psi}$ to the results obtained
within a polynomial expansion of the effective potential
($\left.\frac{m_\sigma}{m_\psi}\right|_{\text{poly}} \approx0.79$ for $\Nf=1$
and $\left.\frac{m_\sigma}{m_\psi}\right|_{\text{poly}} \approx0.786$ for
$\Nf\geq 5$) \cite{Gies:2025pqv}, we observe that the global solution of the
effective potential improves the precision of the mass ratio estimate
specifically for small $N_\text{f}$. This is in line with the observation that
the convergence radius of the polynomial expansion is rather small for small
$N_\text{f}$, as discussed in \cite{Gies:2025pqv}. 

In addition to this quantitative improvement, we emphasize that the global
solution corroborates the mechanism of self-organized criticality induced by the
partial fixed point. The existence of the latter is not merely an artifact of
the polynomial expansion, but rather a genuine feature of the full effective
potential. 

While the present model involves a different set of degrees of freedom, it
remains a remarkable coincidence that the ratio $\frac{m_\sigma}{m_\psi}$ is
quantitatively very close to the Higgs-to-top mass ratio in the Standard Model. 

\section{Conclusions}
\label{sec:conc}

We have revisited the renormalization flow of a relativistic Luttinger-Yukawa
model featuring properties of self-organized criticality in four-dimensional
spacetime. The model can be viewed as the analogue of a $\mathbb{Z}_2$ Higgs-top
model, where the top quark is replaced by a relativistic Luttinger fermion. In
contrast to the Dirac fermionic case, the feature of self-organized criticality
renders all couplings RG irrelevant, and the scalar mass parameter marginal such
that the model exhibits a natural hierarchy of scales and has a large degree of
predictivity. The RG relates self-organized criticality to the existence of a
partial fixed point, which is IR attractive in all couplings within a wide basin
of attraction for generic initial conditions \cite{Gies:2025pqv}.

The present work has extended previous studies of the model in several
directions. Most importantly, we have resolved the full field dependence of the
effective potential, quantitatively improving upon previous estimates based on
local expansions in field space and thereby solving the problem of a small
convergence radius of those expansions. This has allowed us to start from a
functional definition of the partial fixed point and to determine the
corresponding fixed-point potential globally in field space. Our results cover
the large-$N_\text{f}$ limit, and corresponding large- and small-field
expansions fully analytically. Remarkably, the partial fixed-point potential
exhibits a large-field asymptotics familiar from Coleman-Weinberg-type
potentials, which receives mild power-law contributions from the leading
$1/N_\text{f}$ corrections. A full numerical solution for finite $N_\text{f}$ is
presented, and we find close agreement with the analytic results, rather
independently of $N_\text{f}$. For these results, a key ingredient is the
isolation of the only remaining marginal direction of the RG flow at the partial
fixed point, which is associated with the scalar mass parameter. 

Moreover, we have solved the RG flow starting from generic initial conditions in
the symmetric regime, spending a long RG time near the partial fixed point, and
then entering the symmetry-broken regime at naturally small scales. From a
method viewpoint, this has been particularly challenging, as the flow undergoes
several qualitative changes from an initial conventional power-like running, to
a logarithmic running near the partial fixed point, and a rapid transition to
the symmetry-broken regime including the restoration of convexity of the
effective potential. Recent developments in the numerical solution of functional
RG flows have made this possible to a satisfactory degree of accuracy. Our
results for the universal infrared observables in the form of dimensionless
ratios of the vacuum expectation value, the scalar mass, and the fermionic mass
gap improve upon previous estimates specifically for small $N_\text{f}$, where
the convergence radius of polynomial expansions is very small. For large
$N_\text{f}$, our results corroborate the previous estimates. 

For future comparisons with other approaches, it should be noted that the
present model presumably features further perturbatively marginal operators of
four-fermion type, cf. \cite{Gies:2023cnd}, which could be compatible
with the symmetries of the present model and not included in the present study.
If these operators are also controlled by the partial fixed point, they would
not alter the present conclusions. If they contribute marginally, our estimates
for the infrared observables may depend logarithmically on the initial
conditions of these operators. From the examples studied in \cite{Gies:2023cnd},
we expect that such operators also decouple in the large-$N_\text{f}$ limit, and
hence do not affect the present results for large $N_\text{f}$.

Let us finally point out that our quantitative results rely on an improved
leading-order derivative expansion, which leaves room for future systematic
higher-order studies or different expansion schemes. We believe that the
mechanism of self-organized criticality and the underlying partial fixed point
is robust and will persist under such improvements. It remains an interesting
question whether the present model can be fruitfully embedded into a scenario
involving various standard model sectors, where the relativistic Luttinger
fermions become a crucial ingredient for an improved renormalization behavior or
a higher degree of predictivity.  

\section*{Acknowledgments}

We thank Friederike Ihssen, Hannes Keppler, Álvaro Pastor Gutiérrez, Marta
Picciau and Yunxin Ye for valuable discussions. This work has been funded by the
Deutsche Forschungsgemeinschaft (DFG) under Grant Nos. 569836646 (GI 328/13-1)
and 550624556 (Ara3).

\appendix
\setcounter{equation}{0}
\setcounter{figure}{0}
\setcounter{table}{0}
\renewcommand{\theequation}{A\arabic{equation}}
\renewcommand{\thefigure}{A\arabic{figure}}
\setcounter{secnumdepth}{3}

%

\section{Threshold functions}
\label{app:thresh}

\setcounter{equation}{0}

In order to make the paper self-contained, we provide the explicit forms of the
threshold functions used in the main text, for details, see the Appendix of
\cite{Gies:2025pqv}, as well as more general accounts in the literature 
\cite{Jungnickel:1995fp,Berges:2000ew,Hofling:2002hj,Pawlowski:2005xe,Gies:2006wv,Braun:2011pp,Gies:2013pma,Gies:2020xuh}.

The threshold functions are defined in terms of the regularized momentum-space
propagators of fields. We follow the conventions of \cite{Gies:2025pqv} and use
the partially linear regulator (Litim regulator)
\cite{Litim:2000ci,Litim:2001up} for both the scalar and the fermionic sectors.
This allows for an analytic evaluation of the loop momentum integration. The
threshold functions parametrizing these regularized loop integrals then boil
down to simple rational functions of the dimensionless mass parameters,
\begin{align}
& l_{0}^{(\text{L}),d}(\omega ; \eta) = \frac{4}{d} \left(1-\frac{\eta}{d+2} \right) \frac{1}{1+\omega} , \label{eq:lL0l}\\
& l_{1,1}^{(\text{LB}),d}(\omega_{1},\omega_{2};
         \eta_{1},\eta_{2})= \frac{2}{d} \frac{1}{(1+\omega_1)(1+\omega_2)} \nonumber\\
         &\qquad \qquad \times \left(\frac{2}{1+\omega_1} \left(1-\frac{\eta_1}{d+2} \right) \right. \nonumber\\
         &\qquad \qquad + \left.\frac{1}{1+\omega_2} \left(1-\frac{\eta_2}{d+2} \right) \right)  ,       \label{eq:lL11l}\\
& m_{1,2}^{(\text{LB}),d}(\omega_1,\omega_2;\eta_{1},\eta_{2}) 
        = \frac{1}{2} \frac{1}{(1+\omega_1)(1+\omega_2)^2} \nonumber\\
        &\qquad \qquad \times \left( d+1 + \frac{\omega_1-3}{\omega_1 +1} -\eta_2 \right) ,\label{eq:mLB12l}\\ 
& m_4^{(\text{L}),d}(\omega;\eta)= \frac{(1-\omega)^2}{(1+\omega)^4} + \frac{2(d-\eta)}{d-2}  \frac{1-\omega}{(1+\omega)^3} ,  \label{eq:mL4l}\\
&m_2^{(\text{L}),d}(\omega;\eta)= \frac{4}{(1+\omega)^4} ,   \label{eq:mL2l}\\
& l_{0}^{d}(\omega ; \eta) = \frac{2}{d} \left(1- \frac{\eta}{d+2}\right) \frac{1}{1+\omega}, \label{eq:l0}\\
& m_{2,2}^{d}(\omega;\eta) = \frac{1}{(1+\omega)^4}. \label{eq:m22}
\end{align}
Following the conventions of \cite{Gies:2025pqv}, a symmetric smearing has been
used for potential non-analyticities going along with the Litim regulator.

\bibliography{bibliography}


\end{document}